\documentclass[aps,superscriptaddress,eqsecnum,nofootinbib,showpacs,preprintnumbers,twocolumn]{revtex4}
\pdfoutput=1

\usepackage{longtable}
\usepackage{bm}
\usepackage{relsize}
\usepackage{amsfonts}
\usepackage{amsmath}
\usepackage{amssymb,epsf}
\usepackage{latexsym}
\usepackage{graphicx,epsfig}
\usepackage{amssymb}
\usepackage{float}
\usepackage{subfigure}
\usepackage{epstopdf}
\usepackage[colorlinks=true,citecolor=blue,linkcolor=blue,urlcolor=black]{hyperref}
\usepackage{dcolumn}
\usepackage{psfrag}
\usepackage{wrapfig}
\usepackage{makeidx}
\usepackage{epsf}
\usepackage{color}
\usepackage{multirow}
\usepackage{mathtools}
\usepackage{tikz}

\usepackage[normalem]{ulem}

\newcommand{\xmark}{%
\tikz[scale=0.23] {
    \draw[line width=0.7,line cap=round] (0,0) to [bend left=6] (1,1);
    \draw[line width=0.7,line cap=round] (0.2,0.95) to [bend right=3] (0.8,0.05);
}}

\begin{document}

\title{Polarimetry imprints of exotic compact objects: relativistic fluid spheres and gravastars}

\author{Hanna Liis Tamm}
\email{hanna.liis.tamm@ut.ee}
\affiliation{Institute of Physics, University of Tartu, W. Ostwaldi 1, 50411 Tartu, Estonia}

\author{Nicolas Aimar}
\email{ndaimar@fe.up.pt}
\affiliation{Faculdade de Engenharia, Universidade do Porto, s/n, R. Dr. Roberto Frias, 4200-465 Porto, Portugal}
\affiliation{CENTRA, Departamento de Física, Instituto Superior Técnico-IST, Universidade de Lisboa-UL, Avenida Rovisco Pais 1, 1049-001 Lisboa, Portugal}

\author{João Luís Rosa}
\email{joaoluis92@gmail.com}
\affiliation{Departamento de F\'isica Te\'orica, Universidad Complutense de Madrid, E-28040 Madrid, Spain}
\affiliation{Institute of Physics, University of Tartu, W. Ostwaldi 1, 50411 Tartu, Estonia}

\date{\today}

\begin{abstract} 
Recent observations of polarimetric parameters of active galactic nuclei motivate the study of polarization in the spacetime of exotic compact objects which can mimic the features of black holes in the strong field regime of gravity. In this work, we study the properties of two models for ultra-compact objects containing light rings — relativistic fluid spheres and gravastars. We have simulated the orbit of a hot spot around the considered objects in the polarization ray-tracing software GYOTO, and extracted observables, namely integrated images of the Stokes parameters $I, Q, U$; their evolution during the orbit in the $QU$-plane, and the electric vector position angle (EVPA). Our models resemble the black hole qualitatively, with slight additional imprints which provide a useful tool to constrain the spacetime metric of supermassive compact objects with current and future observations, and probe the fundamental properties of these objects. We have found that one fluid star model with a pressure singularity resembles the black hole entirely, while another gravastar features notable differences in the EVPA curve in the low-inclination case. Since differences between the models become more pronounced for a higher inclination, our results could potentially restrict the adequateness of ECO classes with future high-inclination observations.
\end{abstract}


\maketitle

\section{Introduction}\label{sec:intro}

In recent years, the collaborative efforts of the largest experiments in gravitational physics, namely the direct detection of gravitational wave signals arising from the merger of compact object binary systems \cite{LIGOScientific:2016aoc,LIGOScientific:2021djp,KAGRA:2021vkt} by the Ligo-Virgo-Kagra (LVK) collaboration, the observation of shadows of supermassive compact objects in the core of the galaxies M87 and the Milky Way \cite{EventHorizonTelescope:2019dse,EventHorizonTelescope:2021bee,EventHorizonTelescope:2022wkp} by the Event Horizon Telescope (EHT) collaboration, and the observation of infrared flares in orbital motion close to the Innermost Stable Circular Orbit (ISCO) of Sagittarius~A* (Sgr~A*) \cite{GRAVITY:2020lpa,GRAVITY:2023avo} by the GRAVITY collaboration, have provided direct evidence for the existence of compact objects observationally similar to black-holes. These observations are in alignment with the Kerr hypothesis, which states that the final state of a complete gravitational collapse in an astrophysical scenario is well-described by the Kerr metric\cite{Kerr:1963ud,Penrose:1964wq}. This agreement between theoretical predictions and observations in several fronts is considered one of the greatest achievements of Einstein's theory of General Relativity \cite{Will:2014kxa,Yagi:2016jml}.

Despite the apparent success of the black-hole model to explain the experimental results listed above, black-hole spacetimes present problematic physical and mathematical properties, from the existence of a singularity in their interiors \cite{Penrose:1964wq,Penrose:1969pc}, to the violation of the unitary evolution required by quantum mechanics due to the event horizon (EH) \cite{Hawking:1976ra}. To overcome these issues, a wide variety of alternatives to the black-hole model known as Exotic Compact Objects (ECOs) have been developed, some of which reproducing similar observational properties and thus called black-hole mimickers (see Ref. \cite{Cardoso:2019rvt}) for a complete review on the topic.

One of the most popular categories of ECO models are compact objects composed of relativistic fluid components, a category that includes the two prominent families of relativistic fluid spheres \cite{Buchdahl:1959zz,Rosa:2020hex} and gravitational vacuum stars \cite{Mazur:2004fk,Mazur:2001fv,Lobo:2005uf,Visser:2003ge}, also known as gravastars. These two families of models have been extensively studied in terms of their observational properties, and they fulfill several requirements for physical relevance, including linear stability and similar observational imprints to black-hole spacetimes \cite{Cardoso:2015zqa,Rosa:2023hfm,Tamm:2023wvn,Rosa:2024bqv,Chirenti:2007mk,Pani:2009ss,Igata:2025xxb}. These models can be tested with the data acquired by the current and upcoming generation of experiments in gravitational physics \cite{Cardoso:2016oxy,Cardoso:2017cqb,Postnikov:2010yn,Cardoso:2017cfl,Cardoso:2016rao}. 

In recent years, the polarization of light, i.e. the orientation on the sky of the electric field vector component of the electromagnetic wave, emitted from the vicinity of compact objects has been measured by both the EHT~\cite{EHT_2021,EHT_2024} and the GRAVITY~\cite{gravity2018, gravity2023} experiments. The EHT has measured directly the polarization, expressed in terms of the Stokes parameters $I,Q,U,V$, of light coming from the accretion flow of M87* and Sgr~A* allowing the mapping of the magnetic field orientation and degree of order~\cite{EHT2024b}. Such observation allowed for a better constraining of our model and our understanding of astrophysics around compact objects. The GRAVITY collaboration reported in 2023 the detection of 8 flares with linear polarization measurement~\cite{gravity2023}. These data show, similarly to radio data obtained with ALMA~\cite{Wielgus2022}, loops of polarization in the $QU$-plane. These loops are most likely generated by an orbiting hotspot around the compact object. While the nature of the loops are special relativistic effects, their asymmetry is due to general relativistic effects, mostly light bending~\cite{Vincent:2023sbw}. Similar conclusions, i.e. that polarization is sensitive to the space-time curvature, have been made through the study of polarization in photon rings~\cite{Himwich2020,Palumbo2023}. Thus, the study of the polarization of light coming from compact objects appears to be the one of the best methods to probe and diagnose the space-time, and so the nature of compact objects. 

In order to compare the predictions of models of compact objects to observational (photometric) data, one needs to take care of relativistic effects e.g. beaming, relativistic Doppler effect, light bending, and so on. While analytical formulae have been developed for the Kerr geometry, this process is much more difficult for ECO models. To overcome this issue, one can use the ray-tracing method, which consists of integrating the geodesic equation for the photon between an observer and the source of radiation. This method is known as backwards ray-tracing. Nowadays, multiple codes are capable to perform efficient backwards ray-tracing. We chose the public code GYOTO~\cite{Vincent:2011wz,Aimar:2023vcs} for its capability to perform ray-tracing with polarization and its flexibility in implementing other alternative spacetime metrics besides Kerr \cite{Rosa:2022toh,Rosa:2023qcv,Rosa:2025dzq,Macedo:2024qky}. Hot-spot polarization has also been analyzed in other black-hole spacetimes \cite{Angelov:2025rut,Chen:2024jkm,Rosa:2025pqp}.

This work is organized as follows. In Sec. \ref{sec:theory} we introduce the models for relativistic fluid spheres and gravastars to be tested in what follows. In Sec. \ref{sec:polar} we define the polarimetric observables and specify the numerical setup in regards to simulations. Next, in Sec. \ref{sec: results} we analyze the simulated polarimetric signals for both fluid stars and gravastars and explain their origins. Finally, in Sec. \ref{sec:concl} we trace our conclusions and perspectives for future work.

\section{Theory and framework}\label{sec:theory}

In this work we are interested in analyzing the polarimetric signatures of two distinct families of static and spherically symmetric compact objects composed of relativistic perfect fluids with a constant density, namely relativistic fluid stars and gravitational vacuum stars, also known as gravastars. In the following, we adopt the usual set of spherical coordinates $x^\mu=\{t,r,\theta,\phi\}$.

\subsection{Relativistic fluid stars}

Relativistic fluid stars supported by thin-shells \cite{Rosa:2020hex} can be described by a piecewise metric composed of an interior $r<r_\Sigma$ and an exterior $r>r_\Sigma$ regions, where $r_\Sigma$ is the radius of the hypersurface separating the two regions. The interior region is populated by a relativistic fluid with a constant density $\rho_0$ whereas the exterior region is vacuum. The line elements that describe the interior $ds^2_-$ and the exterior $ds^2_+$ regions are given by
\begin{eqnarray}\label{eq:metric_fs_int}
    ds^2_ -&=&-\frac{1}{4}\left(3\sqrt{1-\frac{2M}{R}}-\sqrt{1-\frac{2r^2M}{R^3}}\right)^2dt^2+\\
    &+&\left(1-\frac{2r^2M}{R^3}\right)^{-1}dr^2+r^2\left(d\theta^2+\sin^2\theta d\phi^2\right),\nonumber
\end{eqnarray}
\begin{eqnarray}\label{eq:metric_fs_ext}
ds^2_+&=&-\left(1-\frac{2M}{r}\right)dt^2+\left(1-\frac{2M}{r}\right)^{-1}dr^2+\\
&+&r^2\left(d\theta^2+\sin^2\theta d\phi^2\right),\nonumber
\end{eqnarray}
respectively, where $M$ denotes the total mass of the star, $R\geq r_\Sigma$ is the radius of the star at which the thin-shell is absent, i.e., when the mass of the star is completely distributed in volume. In the inner region where the relativistic fluid lies, the stress-energy tensor $T_{\mu\nu}$ takes the form
\begin{equation}\label{eq:def_tab_fs}
T_\mu^\nu=\text{diag}\left(-\rho_0,p,p,p\right),
\end{equation}
where $\rho_0=\frac{3M}{4\pi R^3}$ is the constant volumetric energy density and $p$ is the isotropic pressure profile of the fluid, which is given as a function of the radius $r$ as
\begin{equation}\label{eq:pressure_fs}
p\left(r\right)=\rho_0\frac{\sqrt{1-\frac{2r^2M}{R^3}}-\sqrt{1-\frac{2M}{R}}}{3\sqrt{1-\frac{2M}{R}}-\sqrt{1-\frac{2r^2M}{R^3}}}.
\end{equation}
It is noteworthy that the central value of the pressure diverges for a star radius $R=R_b\equiv 9M/4$, where $R_b$ is the Buchdahl radius. This implies that the regularity of the model in the smooth case, i.e., in the absence of a thin-shell, is restricted by a maximum compactness. Nevertheless, this limit can be surpassed by choosing $R>R_b$ and $2M<r_\Sigma<R_b$, allowing for solutions with a compactivity arbitrarily close to that of the Schwarzschild BH. 

\subsection{Gravitational vacuum stars}

Similarly to the previous model, thin-shell gravastars \cite{Pani:2009ss} can also be described by a piecewise metric composed of an interior $r<r_\Sigma$ and an exterior $r>r_\Sigma$ regions. The main difference between the two models is that the inner region is populated by an exotic fluid with a constant energy density $\rho_0$ and a negative pressure, whereas the exterior is still vacuum. The line elements that describe the interior $ds^2_-$ and the exterior $ds^2_+$ regions are in this case
\begin{eqnarray}\label{eq:metric_gs_int}
ds^2_-&=&-\alpha\left(1-\frac{2r^2M_\rho}{R^3}\right)dt^2+\left(1-\frac{2r^2M_\rho}{R^3}\right)^{-1}dr^2+\nonumber\\
&+&r^2\left(d\theta^2+\sin^2\theta d\phi^2\right),
\end{eqnarray}
\begin{eqnarray}\label{eq:metric_gs_ext}
ds^2_+&=&-\left(1-\frac{2M}{r}\right)dt^2+\left(1-\frac{2M}{r}\right)^{-1}dr^2+\nonumber\\
&+&r^2\left(d\theta^2+\sin^2\theta d\phi^2\right),
\end{eqnarray}
where $R$ is the radius of the gravastar, $M_\rho$ is the portion of the total mass $M$ that is distributed in the inner volume, and $\alpha$ is a free parameter that controls the mass distribution of the gravastar between the inner volume and the thin-shell. The parameter $\alpha$ can be written as
\begin{equation}
    \alpha=\frac{1-\frac{2M}{R}}{1-\frac{2M_\rho}{R}},
\end{equation}
which implies that if $\alpha=1$ then $M=M_\rho$, i.e., the mass is totally distributed in volume, whereas if $\alpha=1-\frac{2M}{R}\equiv \alpha_{\rm min}$ then $M_\rho=0$ and the mass is totally allocated at the thin-shell. We do not consider $\alpha<\alpha_{\rm min}$ since that provides $\rho_0<0$ which violates the weak energy condition. The stress-energy tensor that describes the fluid in the inner region is thus
\begin{equation}\label{eq:def_tab_gs}
T_\mu^\nu=\text{diag}\left(-\rho_0,p,p,p\right),
\end{equation}
where in this case $\rho_0=\frac{3M_\rho}{4\pi R^3}$ and the pressure $p$ satisfies the equation of state
\begin{equation}\label{eq:pressure_gs}
p=-\rho_0.
\end{equation}
Note that Eq. \eqref{eq:pressure_gs} implies the violation of the strong energy condition, i.e., $\rho+3p<0$, a feature responsible for suppressing the attractive nature of gravity in the inner region. Due to their negative pressure, gravastars are not covered by the Buchdahl theorem and thus they can present compacticities arbitrarily close to that of the Schwarzschild BH independently of their mass distribution.

\section{Polarimetry}\label{sec:polar}

\subsection{Polarimetric observables}

We recur to the ray-tracing software GYOTO \cite{Vincent:2011wz,Aimar:2023vcs} to produce the polarimetric signatures  of the fluid star and gravastar models introduced previously. These signatures are produced through the simulation of the orbit of a hot-spot around a central object described by a fluid star or a gravastar. The software GYOTO generates images containing the simulated values for the specific intensities of the Stokes parameters. In particular, we are interested in the $Q$ and $U$ Stokes parameters, which are defined in terms of the electric field of an incident wave, i.e. the polarization, on the screen of the observer given by \cite{Vincent:2023sbw}
\begin{equation}
    \textbf{E}=E\left(\cos \chi_o \textbf{e}_\alpha+\sin\chi_o \textbf{e}_\beta\right),
\end{equation}
where $E$ denotes the amplitude of the electric field, $\chi_o$ the observed electric vector position angle (EVPA), and the vectors $\left(\textbf{e}_\alpha,\textbf{e}_\beta\right)$ form an orthonormal basis in the screen of the observer. The Stokes parameters $Q$ and $U$ are then defined as
\begin{eqnarray}
    Q&=&I\cos\left(2\chi_o\right),\\
    U&=&I\sin\left(2\chi_o\right),
\end{eqnarray}
where $I=E^2$ is the Stokes parameter $I$ which represents the total intensity of the incident wave. Finally, the observed EVPA is given by
\begin{equation}
    \chi_o=\frac{1}{2}{\rm atan2}\left(U,Q\right),
\end{equation}
which, by definition, falls in the interval $\chi_o\in\left[-\frac{\pi}{2},\frac{\pi}{2}\right]$.

The observed polarization depends mostly on 1) the emission mechanism, here assumed to be synchrotron radiation; 2) the curvature of the geodesics followed by the photons; and 3) the magnetic field configuration. In the rest frame of the emitter, the emitted polarization vector $\mathbf{f_e}$ for synchrotron radiation is orthogonal to both the direction of propagation of the photon $\mathbf{K_e}$ and the magnetic field vector $\mathbf{B_e}$~\citep{RybickiLightman:1979}, i.e.
\begin{equation}\label{eq: f_e}
    \mathbf{f_e} = \mathbf{K_e} \times \mathbf{B_e}.
\end{equation}
The emitted polarization vector in the rest frame of the emitter is expressed by the Stokes parameters following the equations in~\cite{Marszewski:2021}.
By definition, the polarization vector is always, in vacuum, orthogonal to the photon direction, which changes in curved space-time. It must thus be parallel transported from the emitting region to the observer. While it is possible to perform this process analytically in the Kerr space-time~\cite{Gelles:2021}, it is much more difficult for arbitrary space-time and requires geodesic and parallel transport integration, as done in GYOTO (see \cite{Aimar:2023vcs} for more details).  

GYOTO outputs the Stokes parameters in the form of 2-dimensional matrices $S_{lm}^\nu$, where $S=\{I,Q,U\}$, of specific intensities for each instant of time $t_k$. The pixels of the image associated with an observed Stokes parameter correspond to the indices $\{m,l\}$. The simulations are then repeated for different time instants $t_k\in\left[0,T\right[$, where $T$ represents the orbital period of the source, from which one obtains cubes of data $S_{klm}$, where $k$ denotes the temporal index and $l,m$ denote the spatial indices. The time-integrated Stokes parameters are then defined as
\begin{equation}
    \left<S\right>_{lm} = \sum_k S_{klm}.
\end{equation}

In what follows, to compare our results between different fluid spheres and gravastar models and also with the results for the Schwarzschild BH, we consider the following observables: the time-integrated Stokes images $\left<S\right>$, the QU-loops $U\left(Q\right)$, and the temporal EVPA $\chi_o\left(t\right)$.

\subsection{Numerical setup}

Using the GYOTO software, we simulated the orbit of a hot-spot around a central object described by either a fluid star or a gravastar with ADM mass $M=4.2\times 10^6M_\odot$, where $M_\odot$ is the solar mass. We set the distance to the observer at $d=8.25\text{kpc}$, and the emission frequency at $f=230\text{GHz}$
in the reference frame of the observer. We generate images with a resolution of $1000\times 1000$ pixels and a field of view of $250\mu\text{as}$.

The hot spot has a radius of $R_s=0.5M$ and an orbital radius of $r_o=8M$, it orbits at the equatorial plane defined by $\theta=\pi/2$ with Keplerian velocity. The source emits synchrotron radiation from a fiducial thermal distribution of electrons with a number density of $n_e=6.6 \text{cm}^{-3}$, and a temperature of $\Theta_e=200$ in dimensionless units. In accordance with the results of the ALMA \cite{Wielgus2022} and GRAVITY \cite{GRAVITY:2023avo} flare observations, we consider a magnetic field with a vertical orientation and a strength of $B\simeq 0.34\text{G}$, corresponding to a magnetization parameter of $\sigma = 0.01$. The considered models and their relevant features are given in Table \ref{tab:models}, where $R$ denotes the radius of the fluid sphere for FS models, and the radius of the gravastar for GS models. Here, the number density, temperature and magnetic field strength are scaling factors for the flux.

\begin{table}
\begin{tabular}{ c|c|c|c|c } 
 Model & $R/M$& LRs & singularity & EH \\ \hline
 BH & - & 1 & \checkmark & \checkmark \\
 FS1 & 2.25 & 1 & \checkmark & \xmark \\ 
 FS2 & 2.50 & 2 &\xmark & \xmark\\ 
 FS3 & 3.00 & 2 (degen.) & \xmark & \xmark\\
 GS1 & 2.01 & 2 &  \xmark & \xmark\\
 GS2 & 2.50 & 2 & \xmark & \xmark\\
 GS3 & 3.00 & 2 (degen.) &  \xmark & \xmark\\
\end{tabular}
\caption{Spacetime properties of the models considered and in comparison with the Schwarzschild BH, namely the normalized star radius $R/M$ and the presence of (degenerate) LRs, singularities and EHs. We note that the singularities in the BH and FS1 models are of a different nature: the first being a curvature singularity and the second being a pressure singularity. The gravastars have neither singularities nor EHs.}
\label{tab:models}
\end{table}

\section{Results}\label{sec: results}
\subsection{Results for fluid stars}

\subsubsection{Time integrated images}

\begin{figure*}
    \centering
    \includegraphics[width=\linewidth]{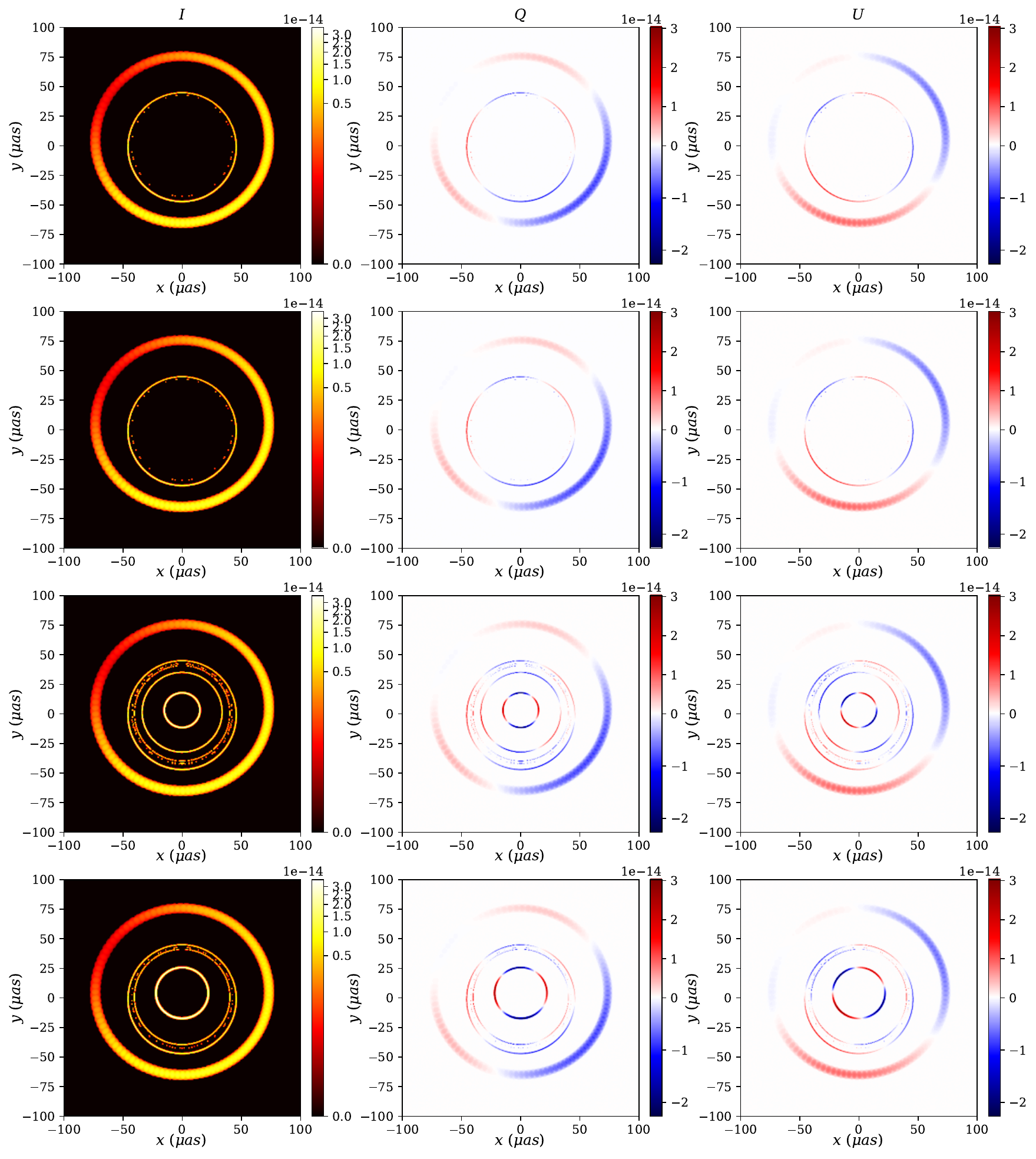}
    \caption{Polarization parameters $I$ (left column), $Q$ (center column) and $U$ (right column) integrated over the hot-spot orbital period for an inclination of $i = 20^\circ$ and hot-spot orbital radius of $r_o = 8M$. We consider the Schwarzschild BH (top row), fluid star models FS1 with $R = 2.25M$ (second row), FS2 with $R = 2.5M$ (third row) and FS3 with $R = 3M$ (fourth row).}
    \label{fig: FS 20}
\end{figure*}

\begin{figure*}
    \centering
    \includegraphics[width=\linewidth]{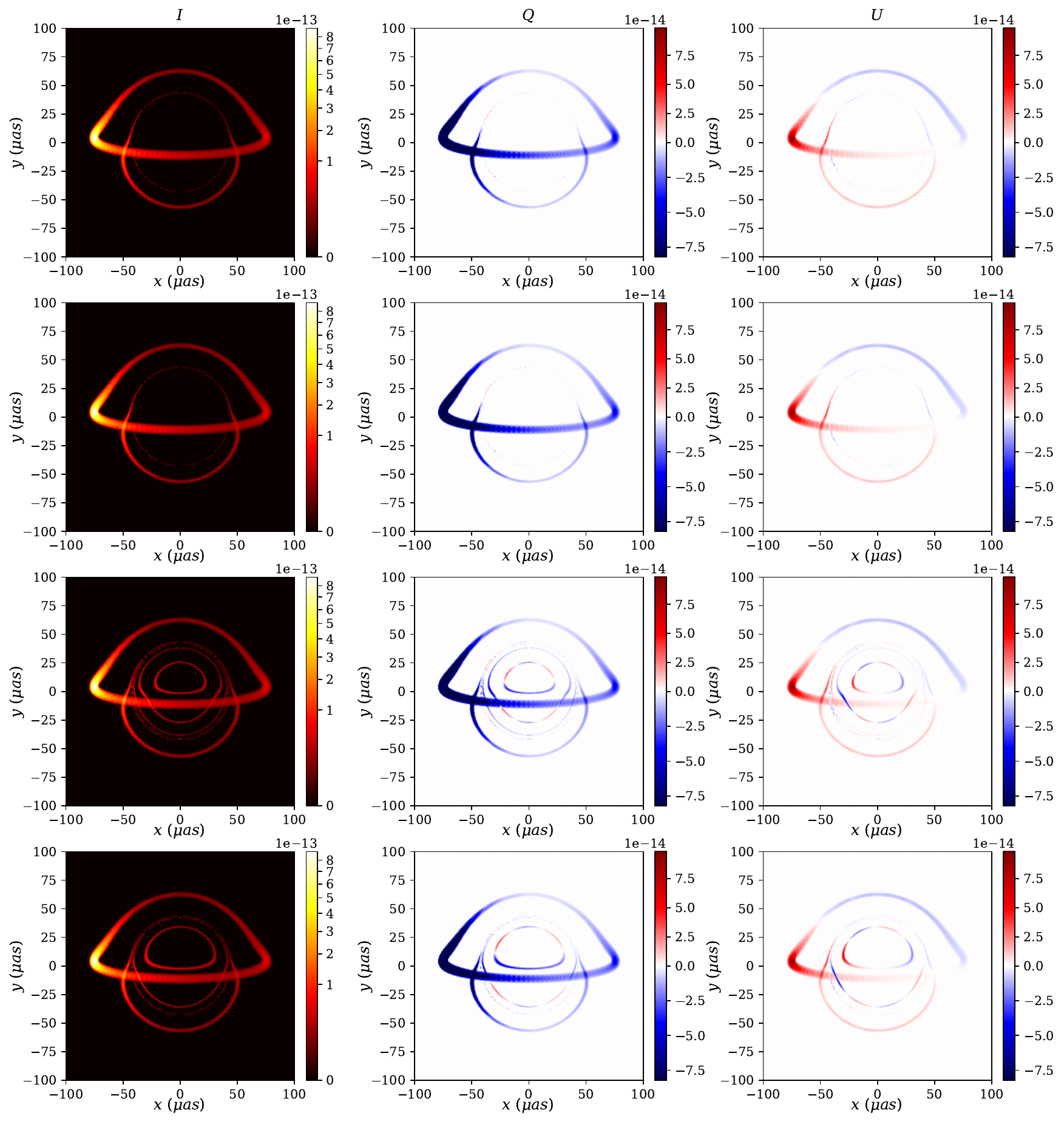}
    \caption{Polarization parameters $I$ (left column), $Q$ (center column) and $U$ (right column) integrated over the hot-spot orbital period for an inclination of $i = 80^\circ$ and hot-spot orbital radius of $r_o = 8M$. We consider the Schwarzschild BH (top row), fluid star models FS1 with $R = 2.25M$ (second row), FS2 with $R= 2.5M$ (third row) and FS3 with $R = 3M$ (fourth row).}
    \label{fig: FS 80}
\end{figure*}

The images generated by the time-integrated stokes parameters I, Q and U for fluid star models with different compactness and for an observation inclination of $20^\circ$ and $80^\circ$ are given in Figs. \ref{fig: FS 20} and \ref{fig: FS 80}, respectively. These results show how different types of relativistic fluid stars and levels of compactness affect the image structures observed.

Consider first the time-integrated images in Figs. \ref{fig: FS 20} and \ref{fig: FS 80}. All models (Schwarzschild BH and FS1-3) feature a wide and uniform outer band, which corresponds to the primary image, for which photons originating from the hot spot go directly toward the observer and cross the equatorial plane only once, corresponding to $n = 0$. The models considered also feature one or more thinner bands with reduced angular width, which represent the secondary images -- photons on these bands make a half-turn around the central compact object with $n = 1$. Finally, we observe discontinuous circles, which we define as light rings (LRs) corresponding to more than two crossings of the equatorial plane ($n \geq 1$). Compared to previous studies on fluid stars, in which emission was assumed to be isotropic, this is nor more the case here. Indeed, synchrotron radiation depends on the angle $\theta_m$ between the direction of the photon and the magnetic field, which here is assumed to be vertical. As a consequence, at low inclination, a novel feature is observed in the form of decreased intensity in the upper left part of the primary image. This originates due to aberration effects caused by the relativistic motion of the hot spot and the direction of its orbit \cite{EHT_2021b} and light bending by the curved space-time.

All considered models are ultracompact, featuring one or more LRs. The model FS1 (second row in Figs. \ref{fig: FS 20}-\ref{fig: FS 80}) resembles the Schwarzschild BH (first row), featuring a primary track, a secondary track, and a single photon ring in the interior. The less compact models FS2 (third row) and FS3 (fourth row) feature multiple secondary, and LR tracks due to the absence of a horizon and, in result, additional polarimetric contributions. For models FS1-3, the signatures in the outer part of the images are identical (outer primary, secondary and LR track), while differences emerge in the inner part of the images. We observe additional LR and secondary tracks as well as an additional plunge-through track with $n =0$ closest to the center.
As the compactness decreases between models FS2-FS3 and the LR becomes degenerate, fewer tracks appear in the interior of the model FS3 than in FS2. Since the intensity of the hot spot is distributed to the interior tracks for the FS models, and interior tracks are obscured by the potential well for the model FS1, the total intensity of this model is reduced with respect to its counterparts FS2-3. At high inclination, the position of the hot spot during the period has a greater effect on the observables. Secondary images and LRs become visible only when the hot spot is passing behind the compact object. Furthermore, the beaming effect becomes stronger as the hot spot approaches the observer, as seen in the increased intensity on the left side of the images.

Consider now the features of the Stokes parameter $Q$, at low inclination. Additional interior images appearing in models FS2-FS3 are mostly symmetric, with negative values dominating in the upper and lower parts of the images and positive values dominating in the left and right portions. However, these images include some skewness in the symmetry -- this represents how the wave vectors of photons on different tracks are affected by the gravitational bending of the FS. For the exterior primary band, we notice two sign changes along the track, whereas for the interior images, the number of sign changes is increased to four. Regarding the integrated Stokes parameter $U$, we observe similarities with the integrated $Q$ parameter. However, in this case, we observe a change of the symmetry angle. Further, in contrast to the parameter $Q$, the outer primary track has four changes of sign for the $U$ parameter. 

Note also that the innermost plunge-through track is the most bright and thick of the interior tracks, and therefore provides the greatest distinction from the BH. Comparing the models FS2 and FS3, we see that since the FS3 metric is less compact, the plunge-through track increases in angular size.

For the integrated images of a higher inclination $i = 80^\circ$ in Fig. \ref{fig: FS 80}, the tracks remain the same qualitatively, but are distorted due to lensing effects. Compared to the low inclination case, however, the outermost primary image and secondary image have strongly negative $Q$ values with no change in signature, whereas the $U$ parameter is in contrast mostly positive with some negative contributions to the primary track. The interior tracks have four sign changes for both $Q$ and $U$ parameters, but the primary changes sign twice during the orbit for the $U$ parameter.

\subsubsection{Time evolution of the polarization}

\begin{figure*}[ht!]
    \centering
    \includegraphics[width=\linewidth]{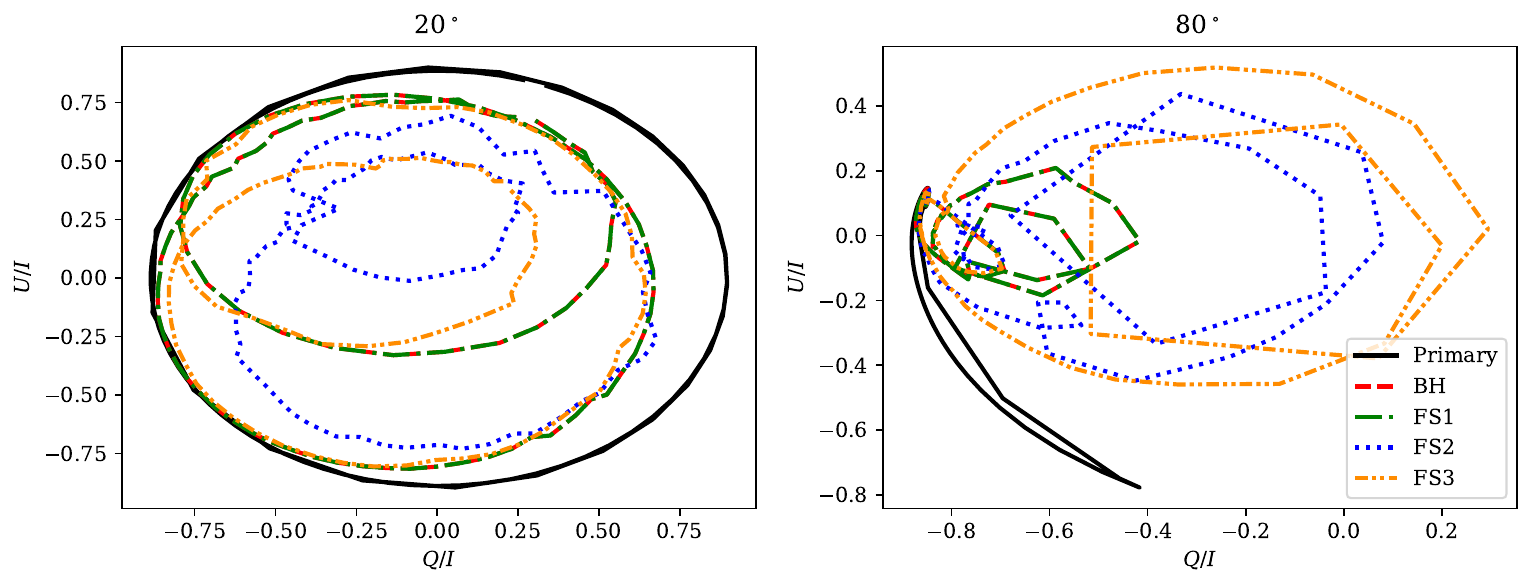}
    \caption{The evolution of the Stokes parameters on the $QU$-plane during the orbital period for the primary image, BH and models FS1-FS3 for inclinations $i = 20^\circ$ (left) and $i = 80^\circ$ (right) at an orbital radius of $r_o = 8M$.}
    \label{fig: FS QU}
\end{figure*}

Let us analyze the time evolution of the image integrated Stokes parameters during the orbital period. The function $U/I(Q/I)$ (which we will call $QU$-loops) for the considered models can be seen in Fig. \ref{fig: FS QU}; for a more detailed plot, see Fig. \ref{fig: QU big} which shows the time evolution more explicitly. Note that to better distinguish contributions from additional images, we have also computed the Stokes parameter values considering only the primary image.

As expected, the $QU$-loop of the primary image for a small inclination $i = 20^\circ$ follows a circle with constant radius during the whole orbit, performing two loops \cite{Vincent:2023sbw}. In the case of BH and FS models, the loops are reduced in size compared to the primary and the second loop performed is smaller than the first loop. The smaller loop represents the part of the orbit when the primary image diminishes and the secondary becomes dominant; since the sign of the interior images opposes the sign of the primary, or between themselves, in general, the total polarization factor is reduced, explaining the smaller second loop. This result indicates an important result -- in the presence of additional images (plunge-through and LRs), the $QU$-loop will have a second smaller loop, assuming that the magnetic field is vertical.

The Schwarzschild and FS1 models are identical in the $QU$-loops -- at the beginning of the orbital period, both $Q/I$ and $U/I$ have total positive values. This corresponds to the image for which the primary image is in the bottom left direction and the secondary image is in the top right direction of the image. As the images rotate counterclockwise during the orbital period, the primary image of $Q$ becomes negative. The parameter $U$ follows the same pattern with a small delay. The second (smaller) loop is similar to the first (large) loop, but additional images reduce its width (polarization fraction).

As the compactness of the fluid star decreases, the contribution of the inner images changes the behaviour of the $QU$-loop. The first, larger loop remains mostly the same size, while the second loop becomes smaller due to interior images opposing the faded primary. The second loop is smallest for model FS2, since it has the largest number of interior tracks that also have opposite signs between themselves, therefore reducing the absolute values of $Q$ and $U$ as the primary image disappears. The model FS3 is more similar to the models with singularities, as the number of interior tracks is smaller with respect to the model FS2, even if it is brighter.

In the case of inclination $i = 80^\circ$ (right image in Fig. \ref{fig: FS QU}), the circular structure becomes less apparent. The loops are skewed towards negative values of $Q/I$, as expected from the integrated image -- only the LR and plunge-through images show positive values. For the image with only the primary track present, the loop is in an extremely distorted crescent shape. For higher order images, the secondary and LR tracks dominate the $Q$ flux at a later time when they are strongly beamed on the left side of the image, and we observe as many as three loops, with the second loop having the greatest width. Contrary to the smaller inclination case, the $QU$-loop increases in width as the FS becomes less compact, signifying an important distinction regarding inclinations. 
As seen from the integrated images, interior tracks include positive values of $Q$ which reduces the dominantly negative contributions toward the positive and thus increase the size of the loops.
Again, the difference between FS2 and FS3 models comes from the number of the additional images and their relative contribution to Stokes $Q$ and $U$. 

We emphasize that the duration of these additional loops is very low (as can be seen from the low number of points for these loops). There are generated when the primary image is low and the additional images suffer strong light bending (upper part of their track) which occurs over a brief period and can be challenging to detect.


\begin{figure*}
    \centering
    \includegraphics[width=\linewidth]{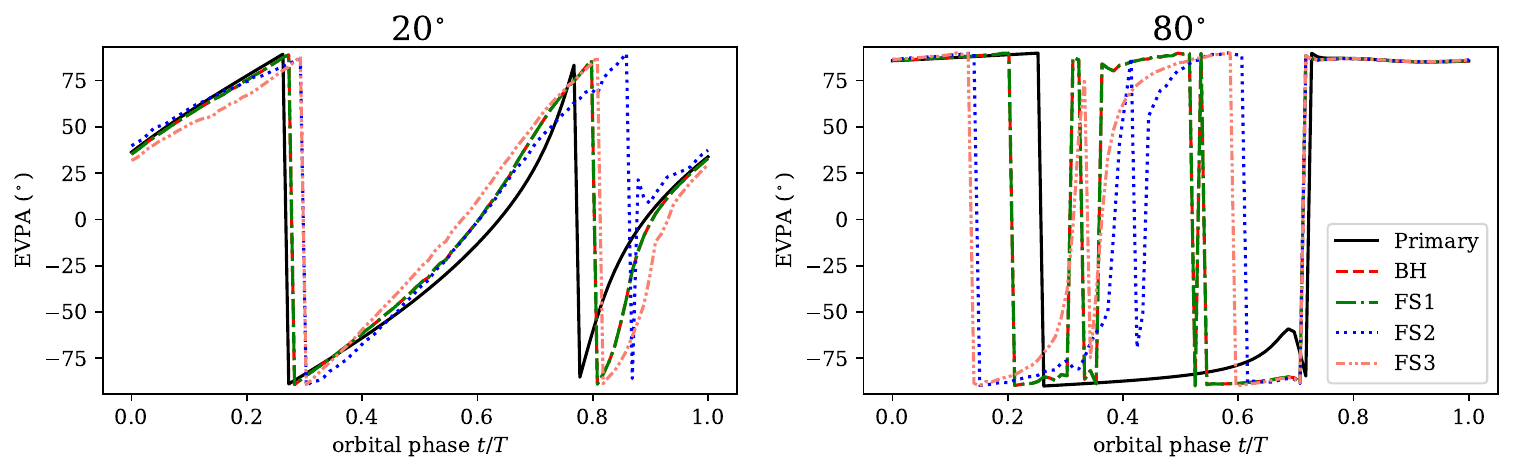}
    \caption{The EVPA evolution during the orbital period for the primary image, BH and models FS1-FS3 for inclinations $i = 20^\circ$ (left) and $i = 80^\circ$ (right) at an orbital radius of $r_o = 8M$.}
    \label{fig: FS EVPA}
\end{figure*}

\subsubsection{Time evolution of the EVPA}

Next, let us consider the EVPA signals of the models in Fig. \ref{fig: FS EVPA} for $20^\circ$ and $80^\circ$. For a lower inclination $i = 20^\circ$, the EVPA mostly increases during the period and has two inversions from $\pi/2$ to $-\pi/2$, representing how the polarization vector of the hot spot slightly changes direction during its orbit. This corresponds to a change of sign of $U$. We observe two notable differences between the models: 1) the time between the inversions and 2) changes in the shape of the EVPA curve, namely additional peaks. These changes appear due to the dimming of the primary image starting near $t \simeq 0.5 T$ and the resulting increased contribution from the interior images.

Again, the model FS1 matches exactly with the BH. The LR and secondary image contributions impose an additional time shift, delaying the inversions compared to the signal with only the primary present. Further, the EVPA increases slightly slower just before the second inversion, increasing the time between the inversions. The next models FS2-FS3 feature an increased time delay and slight noise originating mainly from the LRs. Additionally, at the end of the period, the EVPA decreases for a small interval -- we emphasize that the origin of this feature is not physical, but appears due to the irregularity of the LR tracks due to resolution.

The higher inclination case $i = 80^\circ$ is more complex since the number of inversions is greater. In contrast with the low-inclination case, the EVPA is approximately constant during the whole orbital period due to the strongly negative values of $Q$ with some low-amplitude variation around $\approx 88^\circ$, causing multiple inversions. The inversions are caused by a change of sign of the Stokes parameter $U$ along the orbit as seen from the integrated images. As the simplest example, the primary has two inversions, since there are two changes in sign when the image of the hot spot is in the up left or lower right part of the image (see Fig. \ref{fig: FS 80}). Additional images complicate the structure and increase the number of sign changes and therefore the number of inversions. Similarly to the low inclination case, we notice that the time between the two main inversions of the primary increases with the decrease of the compactness. We notice that the EVPA increases faster at the start of the period since the contribution of the Stokes $U$ of the higher-order images reduce the absolute value of $U \rightarrow 0$ -- thus, less compact models have earlier inversion times. 
The model FS1 and its Schwarzschild counterpart showcase multiple inversions in-between the main inversions, corresponding to small fluctuations near the inversion limit, which are affected by numerical errors and uncertainties. Namely, the inversions are performed when the primary image is dim and the small contributions of LR tracks start dominating. The EVPA is decreased compared to the primary image case, since higher order images altogether oppose the primary image. The model FS2 features numerical inversions as well as smoothly increasing sections which correspond mainly to the evolution of the secondary and plunge-through tracks. We emphasize that these represent physical rather than numerical origins, corresponding to the rotation of the EVPA caused by the increased contribution of higher-order images. We also observe these features -- physical smooth increases and numerical steep increases -- in the least compact configuration FS3.

\subsection{Results for gravastars}

\subsubsection{Time integrated images}
\begin{figure*}
    \centering
    \includegraphics[width=\linewidth]{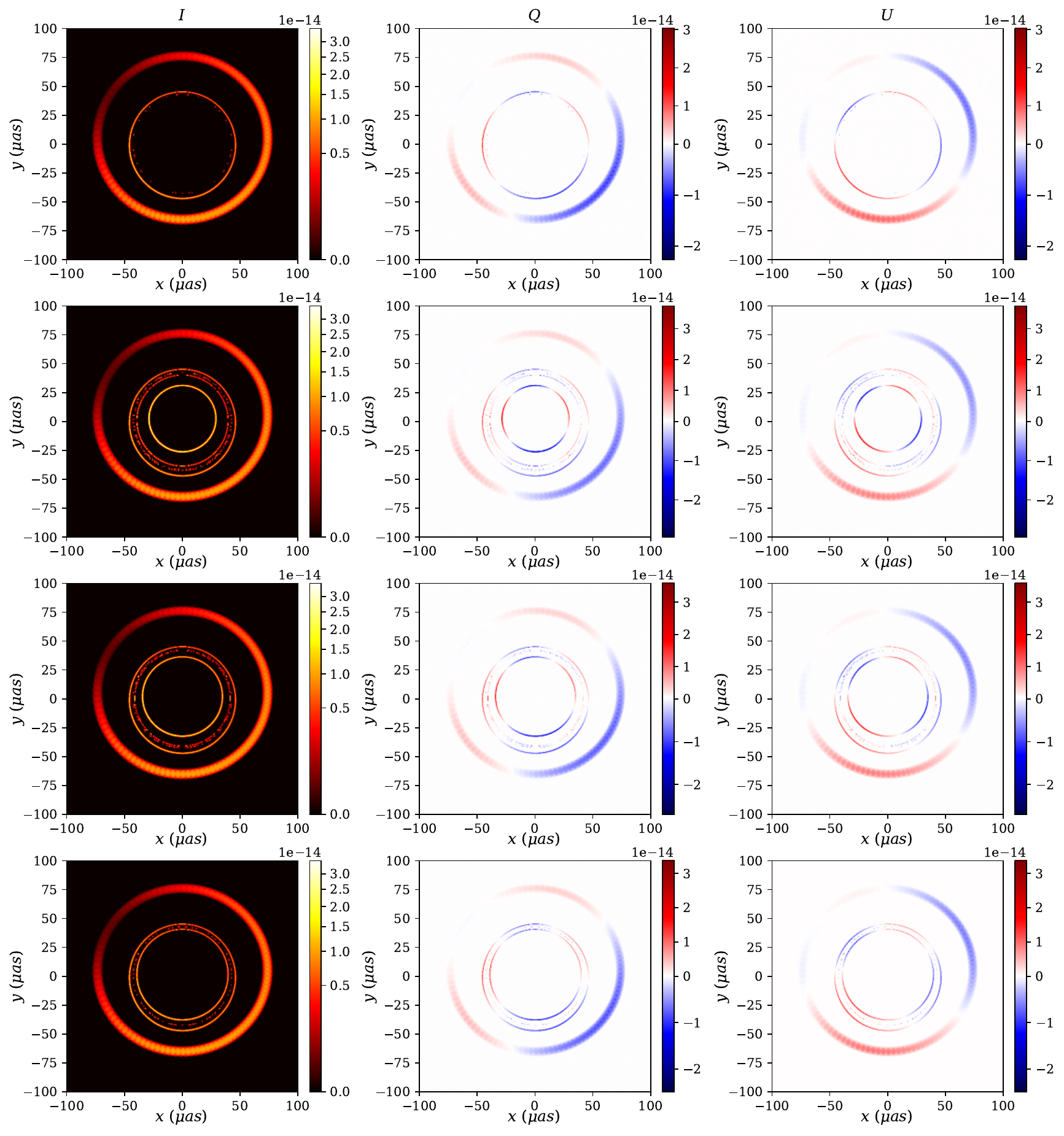}
    \caption{Polarization parameters $I$ (left column), $Q$ (center column) and $U$ (right column) integrated over the hot-spot orbital period for an inclination of $i = 20^\circ$ and hot-spot orbital radius of $r_o = 8M$. We consider the Schwarzschild BH (top row), gravastar models GS1 with $R = 2.01M$ (second row), GS2 with $R = 2.5M$ (third row) and GS3 with $R = 3M$ (fourth row).}
    \label{fig: GS 20}
\end{figure*}

\begin{figure*}
    \centering
    \includegraphics[width=\linewidth]{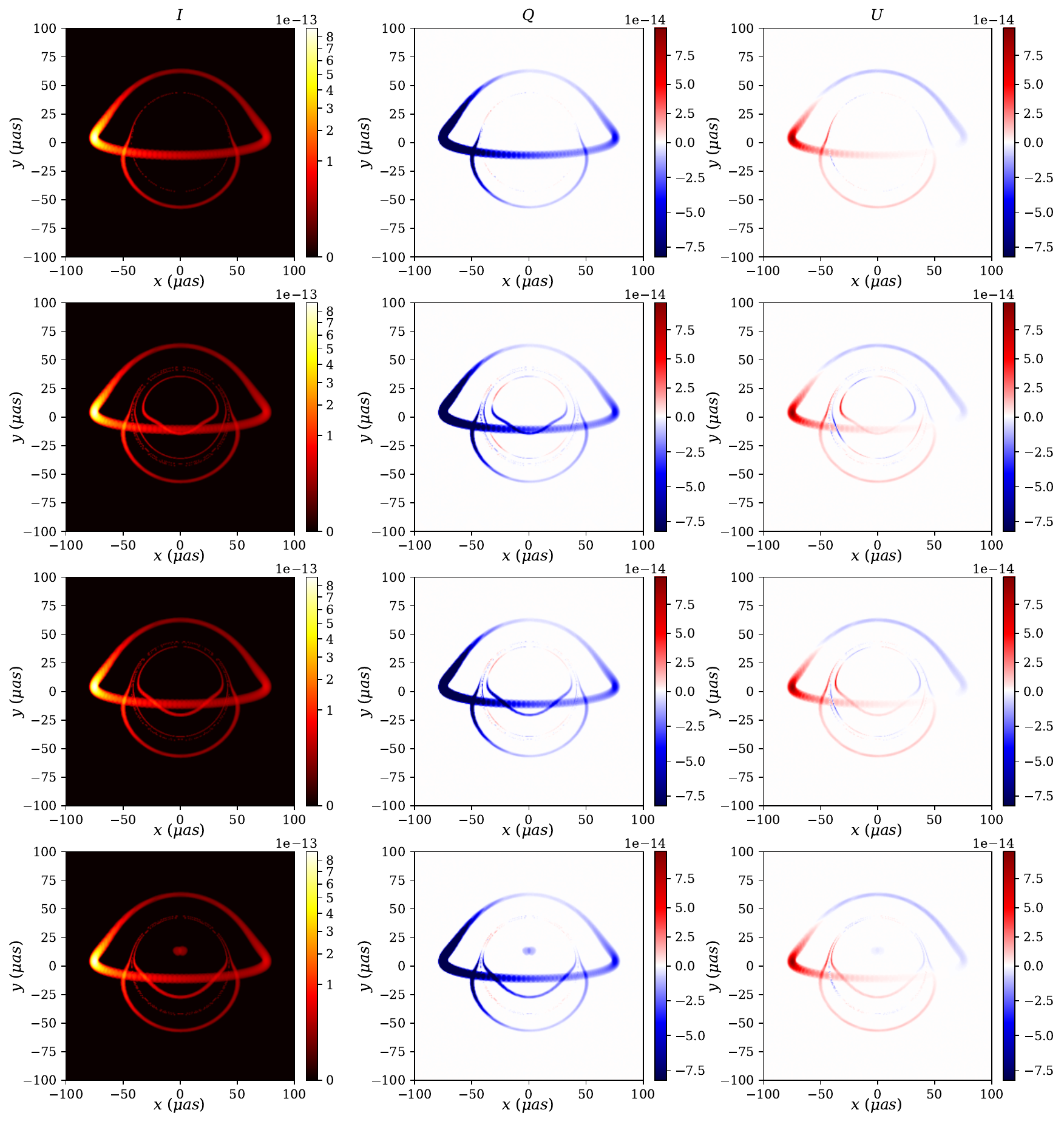}
    \caption{Polarization parameters $I$ (left column), $Q$ (center column) and $U$ (right column) integrated over the hot-spot orbital period for an inclination of $i = 80^\circ$ and hot-spot orbital radius of $r_o = 8M$. We consider the Schwarzschild BH (top row), gravastar models GS1 with $R = 2.01M$ (second row), GS2 with $R = 2.5M$ (third row) and GS3 with $R = 3M$ (fourth row).}
    \label{fig: GS 80}
\end{figure*}

Now let us consider the evolution of the Stokes parameters for gravastars, compared with the BH in the first row of Figs. \ref{fig: GS 20}-\ref{fig: GS 80}. The models GS1-GS3 (in rows 2-4) all show additional images in the interior in addition to the single BH primary, secondary and LR tracks. Contrary to FSs, however, the most compact gravastar differs significantly from the classical case. 
As the compactness decreases, the amount of additional tracks is smaller and the tracks have greater angular width. Similarly to FS models, the GS models exhibit primary, LR and secondary tracks in the interior with symmetry in the up and down directions for the Stokes parameter $Q$, and diagonal symmetry for the Stokes parameter $U$. In addition, we observe dimming in the northwest direction of the image imposed by the vertical magnetic field configuration.

Higher inclination images in Fig. \ref{fig: GS 80} show similar features to FS models, with the exception of the inner track situated further from the center for models GS1-GS2. Regarding the model GS3, we observe a much more narrow inner plunge-through track, reducing its relative contribution compared to the FS3 plunge-through track.

\subsubsection{Time evolution of the polarization}
\begin{figure*}
    \centering
    \includegraphics[width=\linewidth]{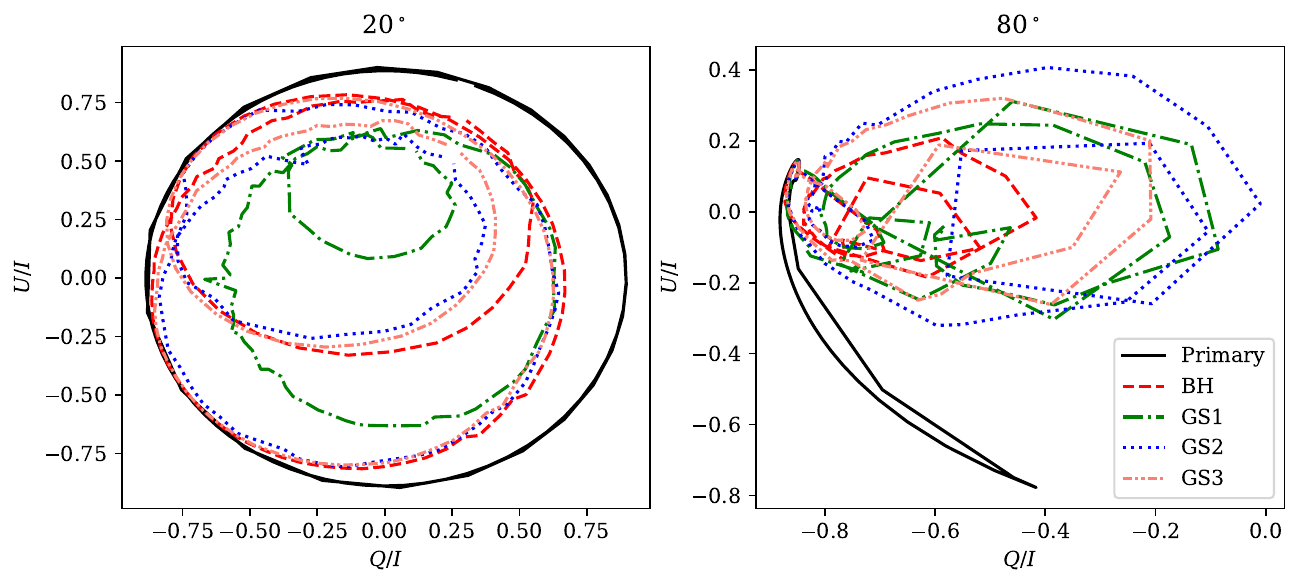}
    \caption{The evolution of the Stokes parameters on the $QU$-plane during the orbital period for the primary image, BH and models GS1-GS3 for inclinations $i = 20^\circ$ (left) and $i = 80^\circ$ (right) at an orbital radius of $r_o = 8M$.}
    \label{fig: GS QU}
\end{figure*}

\begin{figure*}
    \centering
    \includegraphics[width=\linewidth]{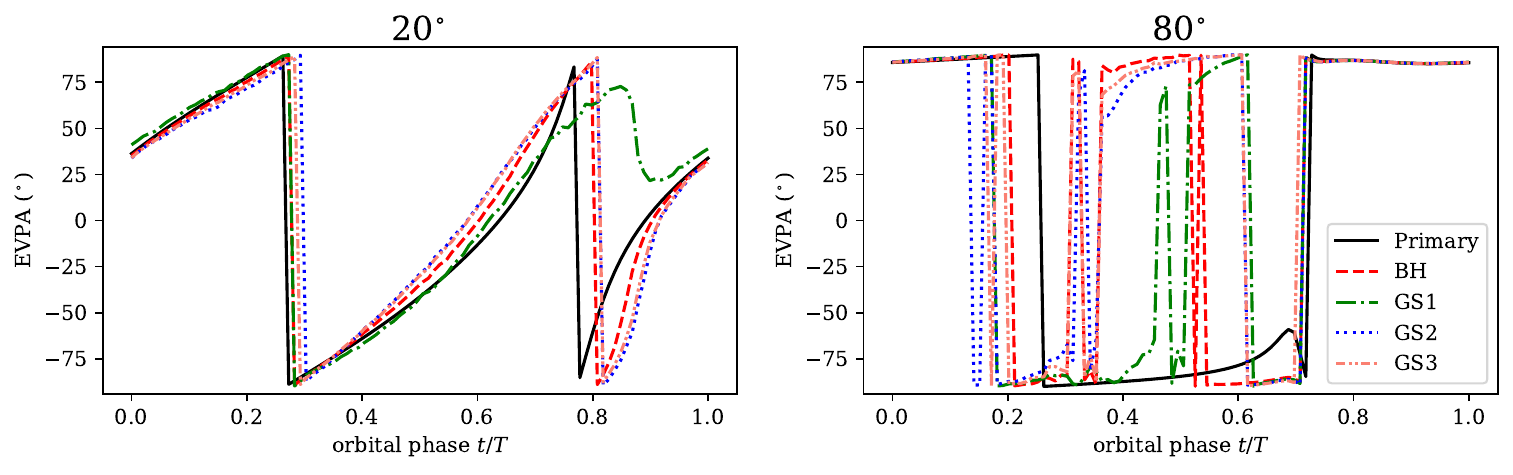}
    \caption{The EVPA evolution during the orbital period for the primary image, BH and models GS1-GS3 for inclinations $i = 20^\circ$ (left) and $i = 80^\circ$ (right) at an orbital radius of $r_o = 8M$.}
    \label{fig: GS EVPA}
\end{figure*}

The $QU$-loops of gravastars GS1-GS3 in addition to the BH and primary configurations are given in Fig. \ref{fig: GS QU} and the EVPA in Fig. \ref{fig: GS EVPA}. We have also included the time-dependent $QU$-loops in Fig. \ref{fig: QU big}.

In a close-to-face-on view, the $QU$-loops of GS configurations mimic the Schwarzschild model, with the loops having similar angular widths. Note that the loop of the most compact model GS1 differs the most from the BH, having the narrowest width -- this feature can be explained by the integrated images in Fig. \ref{fig: GS 20}, where the additional interior images which have different sign for the $Q$ and $U$ parameters compete with each other during the dimming of the primary image resulting in a decrease of the polarization fraction in this part of the orbit.

While the less compact configurations GS2-GS3 have smaller $QU$-loops than the BH model, their loops still exceed the GS1 loop in width due to fewer interior tracks and lower values of intensity of the polarization parameters. In the case of higher inclination, this feature becomes less important and the loops are mostly the same size, with the GS2 model having the greatest width and the GS1, GS3 configurations proving difficult to distinguish.
We can, however, distinguish the amount of loops performed for edge-on views (see also Fig. \ref{fig: QU big} for a clearer distinction). The primary has a singular loop, as seen before; the BH and GS2, GS3 models have three loops with the first and last smaller loops performed respectively during the beginning and end of the period; but the GS1 model differs from the previous models, having as many as five loops. While the origin of these loops is unclear due to an insufficient time sampling and resolution, we can draw a general solution -- higher-order images contribute to the total polarization in a significant way and produce potentially detectable differences.

For a comparative look at the EVPA signatures for gravastars, the change in the EVPA for $i = 20^\circ$ is close to the Schwarzschild curve with the notable exception of model GS1. Models GS2-GS3 differ slightly from the BH because of the longer time delays between inversions -- similarly to before, this feature arises due to a slower increase of the EVPA function. Quantitatively, the change in the time delay is smaller than the delay for fluid stars. The most compact model GS1 behaves differently near the second inversion -- since for this model, the polarization parameters have the smallest values (due to competition between the various images), the EVPA increases even more slowly, thus not reaching an inversion, and even decreases before returning to its original value.

The case for higher inclinations features many inversions, but similarly to fluid stars, we observe regions of smooth increase of the EVPA function, implying a physical origin. These appear slightly earlier for the GS2 and GS3 models, but later for the GS1 model. Additionally, we observe that the smooth increase is steeper for more compact gravastar models, since the appearance of the interior secondary image decreases the EVPA compared to the primary.
\section{Conclusions}\label{sec:concl}

In this work, we have simulated and analyzed the polarimetric signatures of a hot spot orbiting fluid stars and gravastars, including the evolution of the Stokes parameters, $QU$-loops and EVPA during the orbital period. Our analysis has investigated the following ultracompact fluid star models: FS1, which has a radius equal to the Buchdahl limit $R = 2.25M$; FS2 with a radius of $R = 2.5M$, which features a pair of non-degenerate LRs; FS3 with $R = 3M$ which marks the limit between ultra- and non-ultracompactness and has a pair of degenerate LRs. We have also investigated ultracompact gravastar models such as GS1 with a radius equal to $R = 2.01M$, which approaches the limit of the most compact gravastar possible and has two non-degenerate LRs; GS2 with $R = 2.5 M$ which similarly features a non-degenerate LR pair; and GS3 with $R = 3M$ having a degenerate LR pair.

We have found that the polarimetric observables of the investigated models differ from the classic Schwarzschild BH case in many aspects, with the notable exception of the first fluid star model. Indeed, FS1 reproduces exactly the results of the Schwarzschild BH, indicating that the effects of a pressure singularity and the consequent central potential well in these configurations produce the same observational features as those of a curvature singularity and an event horizon. The differences in the other FS models arise from the absence of horizons or potential wells in these models, adding interior tracks to the time-integrated images which can change the total polarization factor. Regarding $QU$-loops, we have observed the expected outcome for a vertical magnetic field, namely two loops for low inclinations. However, these loops vary in size, depending on the compactness of the configuration because of the change of polarization fraction from the interior images. We again emphasize the important result that current observations are consistent with images for which at least a secondary track is present. The EVPA curves as a function of time differ from the BH case, exhibiting a time delay between inversions and modifying the shape of the curve during the second part of the orbital period, when additional images start to dominate.
More differences appear in the high-inclination case, increasing the number of inversions and shifting in addition to steepening the curve for smooth sections.

Our results imply that fluid stars and gravastars have features similar and different of BHs, possibly predicting results obtained in future observations. Although one fluid star model FS1 mimicks the BH exactly in all observables, its pressure singularity deems it as an unphysical solution to the problem, but simultaneously showcases the potential of horizonless alternatives to reproduce the observables of the BH. We have also found a gravastar model GS1 which differs regarding the EVPA, having a single inversion for a small inclination. In conclusion, while it is possible that the differences in low-inclination cases
are not able to constrain the metric in current observations (with the exception of model GS1), the results of this work may prove to be a useful tool in future high-inclination observations. 


\begin{acknowledgments}
J. L. Rosa is supported by the Project PID2022-138607NBI00, funded by MICIU/AEI/10.13039/501100011033 (``ERDF A way of making Europe" and ``PGC Generaci\'on de Conocimiento").
\end{acknowledgments}
\pagebreak
\section*{Appendix}
\begin{figure*}
    \centering
    \subfigure{\includegraphics[width = .49\linewidth]{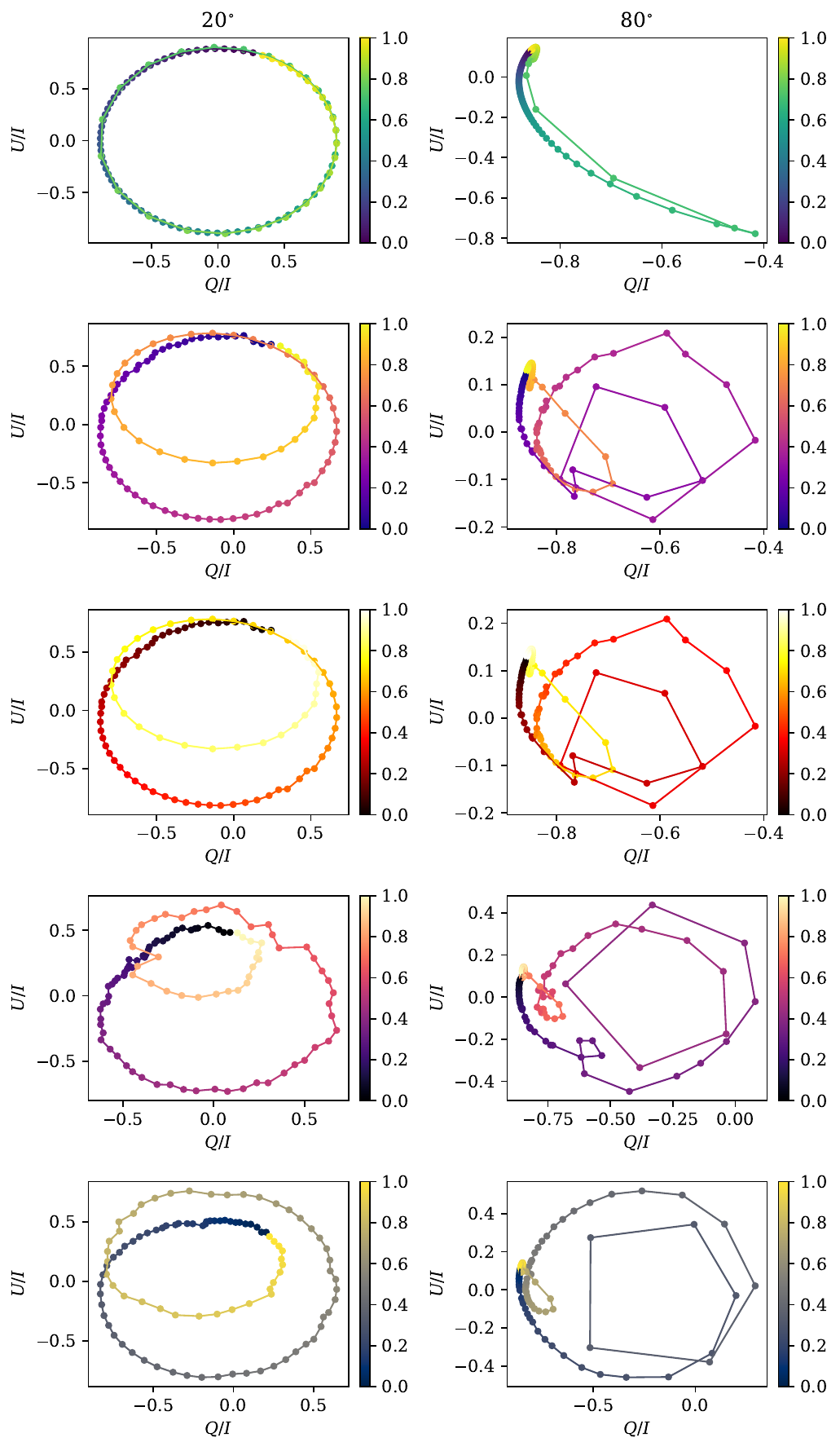}}
    ~
    \subfigure{\includegraphics[width = .49\linewidth]{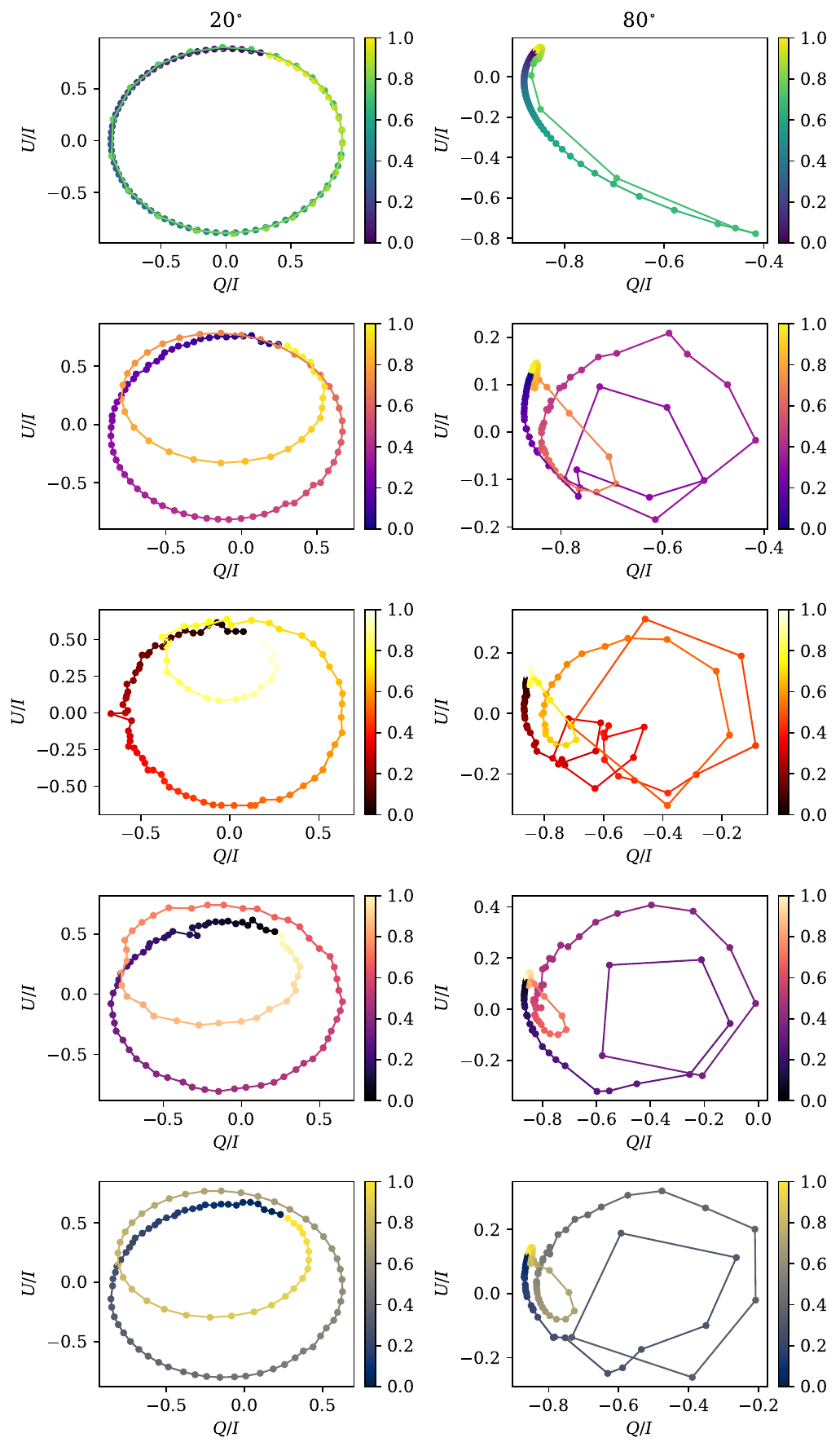}}

    \caption{The $QU$-loops with specific time points for models including the primary (first row), BH (second row), and FS1/GS1 (third row), FS2/GS2 (fourth row) and FS3/GS3 (fifth row) on the left/right figure. The colorbar denotes the orbital time ranging from $0-1 (t/T)$.}
    \label{fig: QU big}
\end{figure*}

To clarify the time evolution of the $QU$-loops, we have plotted the $QU$-loops for FS and GS models with an explicit time dependence in Fig. \ref{fig: QU big}.

\end{document}